\documentclass[prb,twocolumn,showpacs,preprintnumbers,amsmath,amssymb]{revtex4}
\usepackage{graphicx,lscape}
\usepackage{dcolumn}
\usepackage{bm}
\begin{document}
\title{
 Relation between inelastic electron tunneling and vibrational excitation
 of single adsorbates on metal surfaces
 }
\author{
S.~G.~Tikhodeev$^{1,2}$
and H. Ueba$^{2}$
}
\affiliation{
$^{1}$  A.M. Prokhorov General Physics Institute RAS, Vavilova 38, Moscow 119991, Russia \\
$^{2}$ Department of Electronics, Toyama University, Gofuku, Toyama, 930-8555, Japan
}
\begin{abstract}
We analyse theoretically a relation between the vibrational generation rate
of a single adsorbate by tunneling electrons and the inelastic tunneling (IET)
current in scanning tunneling microscope, and the influence of the
vibrational excitations on the rate of adsorbate motions.
Special attention is paid to the effects of finite lifetime of the
vibrational excitations. We show that in the vicinity and below the
IET threshold the rate of adsorbate motion
deviates from a simple power-law dependence on the bias voltage
due to
the effects of bath temperature and adsorbate vibrational lifetime broadenings.
The temperature broadening appears to be confined near the threshold voltage
within a narrow region of several $k_B T$, whereas the lifetime broadening manifests
itself in a much wider region of applied voltages below the IET threshold.
\end{abstract}

\maketitle

\section{INTRODUCTION}
Since a pioneering demonstration of a single atomic switch by
Eigler \textit{et al.},\cite{Eigler91} in which high and low current
state are realized via controlling  Xe atom transfer between a Ni
substrate and a tip of scanning tunneling microscope (STM), there
have been very exciting developments in the unique ability of STM
to induce the motions and reactions of single adsorbates at
surfaces. Examples of such novel experiments are dissociation of
decaborane on Si(111),\cite{Dujardin92} desorption of hydrogen from
hydrogen-terminated Si(100),\cite{Shen95} a step-by-step control of
chemical reaction to form a biphenyl molecule from a iodobenzene
on Cu(111),\cite{Hla00} rotation and dissociation of an oxygen
molecule on Pt(111),\cite{Stipe97} rotation of acetylene
molecule on Cu (100),\cite{Stipe98}  hopping of CO on Pd(110),\cite{Komeda}
and hopping and desorption of ammonia on Cu(100).\cite{Pascual03}

The theoretical understanding of the physical mechanisms behind these
motions and reactions of single adsorbates has made a slow but steady
progress toward a full understanding of how tunneling electrons couple to
nuclear motion of an adsorbate to overcome the potential barrier along
the relevant reaction coordinate. The power-law dependence of the atom
transfer rate as a function of applied voltage or tunneling current
observed for the Eigler switch has been modelled as a potential barrier
crossing between the potential wells formed by the interaction of the the
adatom with the tip and the substrate, respectively. The atom overcomes
the potential barrier through the stepwise vibrational ladder climbing by
inelastic tunneling electrons. The so-called vibrational heating due to
inelastic tunneling electrons has been proposed  by Gao
\textit{et al.},\cite{Gao92} Walkup \textit{et al.},\cite{Walkup93} and
Brandbyge \& Hedegard \cite{Brandbyge94}
[see also a detailed discussion in Ref.~\onlinecite{Gao97}].
Arrhenius-like expression for
the rate of motion is characterized by an \textit{effective} temperature
$T_\mathrm{eff}$ in the presence of inelastic tunneling current, otherwise a
vibrational mode is in thermal equilibrium  with a substrate temperature
${T}$.

Another possibility of \textit{coherent} multistep vibrational
excitation by a single electron has been proposed by Salam
\textit{ et al.},\cite{Salam94} in analogy to a mechanism of
desorption induced by electronic transition  developed by
Gadzuk.~\cite{Gadzuk91} The coherent multiple excitation mechanism
was shown to be important at low tunneling currents, where the
average time between successive electrons is longer than the
vibrational lifetime. Dissociation rate of single O$_2$ on
Pt(111), where the vibrational relaxation rate of the O-O stretch
mode due to electron-hole pair excitations in the substrate is
much larger than the tunneling current,\cite{Persson87}  has been
satisfactory described by this coherent
multiple excitation process.\cite{Stipe97}
However, one should bear in mind
that since both coherent and incoherent multiple excitation
mechanisms show a power-law dependence over a limited range of the
bias voltage or the tunneling current, they can not be simply
distinguished.

Recently we have developed~\cite{Tikhodeev01,Mii02,Mii03} a theory of vibrational tunneling
spectroscopy of adsorbates on metal surfaces using the
Keldysh-Green's function method~\cite{Keldysh,Arseyev92} for an adsorbate-induced resonance
model proposed by Persson and Baratoff.\cite{Persson87-02} This has
enabled us to elucidate the elementary processes of electron
transport via adsorbate state, and to obtain general formulae for
the elastic and inelastic electron tunneling (IET), and the
effective vibrational distribution function within the
second-order perturbation over the electron-phonon interaction of
adsorbates.
In comparison with the earlier papers,\cite{Gao92,Brandbyge94}
where the Keldysh formalism was used as well, we take into account
self-consistently the effects of the finite vibration lifetime
on the IET spectra as well as on the vibration generation rate.

In this work a relation between the
IET current and vibrational generation rate is studied using the coupled
Dyson and kinetic equations  for causal and statistical
phonon Keldysh-Green's functions. The stationary
nonequilibrium phonon distribution function characterized by the
current-driven vibrational generation rate and the vibrational
relaxation rate allows us to introduce the effective temperature
and to describe self-consistently the IET-induced vibrational
heating. A primary interest is focused on
the vicinity and below the threshold of the vibrational excitation.
This situation was not investigated
in Refs.~\onlinecite{Gao92,Walkup93,Brandbyge94,Gao97}, where
the limit of large applied voltage was considered, $eV \gg \Omega$ ($\Omega$ is the
vibrational mode energy).
As we show, the dependencies of the vibrational generation rate
and of the multi-step adsorbate excitation rates
on the bias voltage deviate from simple power laws due to
the effects of bath temperature and a finite lifetime of the vibrational
mode excited by tunneling electrons.
Whereas the bath temperature broadening is restricted to a
narrow region within several $k_B T$, the vibrational lifetime broadening
causes a strong overheating effect in a much wider bias voltage region
below the IET threshold.

The paper is organized as follows. In Section \ref{sec:theory} the general
formulation of the problem in terms of Keldysh-Green's functions is outlined,
and the kinetic equation for vibrational excitation is derived. In Section \ref{sec:Phgenrate}
we analyze the expressions for the IET vibrational generation rates and IET current.
The physical meaning of IET vibrational generation rate is discussed in Subsection
\ref{subsec:Giet}. In subsection \ref{subsec:Teq0} the limiting cases of low bath
temperature and no vibrational broadening are analyzed, and
it is shown that in this limit our results coincide with those received previously
in Refs.~\onlinecite{Gao92,Walkup93,Brandbyge94,Gao97}. Numerical examples are
discussed in Subsection \ref{subsec:Num}.

\section{Theory}\label{sec:theory}
We consider the same Hamiltonian as used before~\cite{Persson87-02,Gao97,Tikhodeev01,Mii02,Mii03}
\begin{eqnarray}
{\cal H} &=& \sum_k \varepsilon_k c^{\dagger}_k c_k +
\sum_p \varepsilon_p c^{\dagger}_p c_p +
\varepsilon_a c^{\dagger}_a c_a +
\Omega (b^{\dagger} b + 1/2) \nonumber\\
&+&
\sum_k \left( V_{k a}c^{\dagger}_k c_a + {\rm h. c.}\right) +
\sum_p \left( V_{p a}c^{\dagger}_p c_a + {\rm h. c.}\right)
\nonumber \\
& + &
\chi c^{\dagger}_a c_a(b^{\dagger} + b),
\label{eq:Hamiltonian}
\end{eqnarray}
where the energies and annihilation operators of a substrate, a
tip, an adsorbate orbital and a vibrational mode are denoted by
$\varepsilon_p, \varepsilon_k, \varepsilon_a, \Omega$ and
$c_p, c_k, c_a, b$, respectively. The tip and substrate systems
are assumed to be in thermal equilibrium at the same temperature
$T$, and to have independent chemical potentials $\mu_t$ and
$\mu_s$, respectively, whose difference corresponds to the
bias voltage $eV = \mu_s - \mu_t$. Electronic
tunneling matrix elements $V_{ka}$ (tip-adsorbate) and
$V_{pa}$ (substrate-adsorbate) give rise to stationary current
between the tip and the substrate through the adsorbate orbital.
 The distribution function for electrons of the tip and substrate systems,
$n_t$ and $n_s$ respectively,
are given by the Fermi distribution function. $\chi$ is the vibration-adsorbate orbital
coupling constant.

In the Keldysh-Green's function method~\cite{Keldysh} the nonequilibrium process of
the electron tunneling between the tip and substrate through the adsorbate level is described
by the \textit{coupled} Dyson and kinetic equations for causal and statistical Keldysh-Green's functions.
This \textit{coupled} description of the dynamical and statistical properties of the nonequilibrium
system allows, within the Keldysh-Green's function method, to find \textit{self-consistently}
the spectra \textit{and} occupation numbers changes which are introduced into the system
by the external action (the tip-substrate bias voltage in our case), mediated by the
interactions in the system. In what follows, the implementation of the
the Keldysh-Green's function method for the Hamiltonian~(\ref{eq:Hamiltonian}) is outlined
by Eqs.~(\ref{D+-2})-(\ref{Na}) without detailed explanations;
the latter can be found in the original paper
by Keldysh~\cite{Keldysh}, see also in Refs.~\onlinecite{Arseyev92},~\onlinecite{Tikhodeev01,Mii02,Mii03}.

The kinetic equation for vibrational excitations (phonons) takes the form
\begin{equation}\label{D+-2}
 \frac{\partial N_\mathrm{ph}}{\partial t}=\int \frac{d\omega}{2\pi}\bigl[
    \Pi^{+-}(\omega)D^{-+}(\omega)-
    \Pi^{-+}(\omega)D^{+-}(\omega)
    \bigl].
    \end{equation}
Here the phonon Keldysh-Green's functions $D^{\pm \mp}$ are given by
\begin{eqnarray} \label{D+-}
     D^{+-}(\omega) & =& n_\mathrm{ph}(\omega)[D^r(\omega)-D^a(\omega)]
     \\ \nonumber
     &=&
     -2i \pi n_\mathrm{ph}(\omega) \rho_\mathrm{ph}(\omega),\\
      \\ \nonumber
      \label{D-+}
      D^{-+}(\omega)&=&[1+ n_\mathrm{ph}(\omega)][D^r(\omega)-D^a(\omega)]
     \\ \nonumber
     &=&
-2i \pi [1+ n_\mathrm{ph}(\omega)]\rho_\mathrm{ph}(\omega),
\end{eqnarray}
where
$n_\mathrm{ph}(\omega)$ is the vibration  occupation function to be determined
self-consistently from the kinetic equation, and the phonon  retarded and advanced
Green's functions are given by
\begin{eqnarray} \label{Dra}
D^r(\omega)=\frac{1}{\omega-\Omega+i\gamma_{eh}(\omega)/2}, \\ \nonumber
D^a(\omega)=\frac{1}{\omega-\Omega-i\gamma_{eh}(\omega)/2}.
\end{eqnarray}
Here $\gamma_{eh}(\omega) = -2\,\mathrm{Im}\,\Pi^r $  (where $\Pi^r $ is
the retarded phonon polarization operator)
is the vibrational damping rate due to electron-hole pair excitations
in the substrate and tip.\cite{footnote} We omit here the real part of $\Pi^r $,
which gives a red shift of the resonant frequency $\Omega$.

The phonon polarization operators
$\Pi^{\pm \mp}$ and $\Pi^r$ are calculated
in the second order perturbation over $\chi$.
For example,
\begin{equation}\label{Pi+-}
 \Pi^{+-} = i \chi^2 \int \frac{d \varepsilon}{2 \pi} G_a^{+-}(\varepsilon + \omega)
 G_a^{-+}(\varepsilon),
\end{equation}
\begin{equation}\label{Pir}
 \Pi^r = i \frac{\chi^2}{2} \int \frac{d \varepsilon}{2 \pi}
  \left[ G_a^a(\varepsilon) F_a(\varepsilon + \omega) + F_a(\varepsilon) G_a^r(\varepsilon + \omega)
  \right],
\end{equation}
where the retarded, advanced and statistical Keldysh-Green's functions of the adsorbate
are connected as
\begin{eqnarray}\label{FGr}
F_a(\varepsilon) &=& [1 - 2 n_a(\varepsilon)][G_a^r(\varepsilon)-G_a^a(\varepsilon)]
\\ \nonumber
&=& - 2 i \pi [1 - 2 n_a(\varepsilon)] \rho_a(\varepsilon),
\end{eqnarray}
with the adsorbate density of states $\rho_a$ and occupation function $n_a$,
\begin{equation}\label{rhoph}
\rho_a(\varepsilon) = (1/\pi)\Delta/
[(\varepsilon - \varepsilon_a)^2+ \Delta^2],
\end{equation}
\begin{equation}\label{Na}
     n_{a}(\varepsilon) = \frac{
    n_{s}(\varepsilon) \Delta_{s} + n_{t}(\varepsilon) \Delta_{t}
}{\Delta}\, .
    \end{equation}
Here ${\mit \Delta}={\mit \Delta}_{t} + {\mit \Delta}_{s}$
is the resonant width of the adsorbate orbital,
${\mit \Delta}_{\rm t(s)}(\varepsilon)=\pi \sum_{k(p)}|V_{k(p)a}|^2 \delta(\varepsilon - \varepsilon_k)$
is the partial width due to the hybridizations
between the tip and the adsorbate (substrate and adsorbate).
The corrections (due to the vibration-adsorbate interaction)
to the adsorbate density of states and to the adsorbate
occupation function, Eqs.(\ref{rhoph}) and (\ref{Na}), respectively, are in the
second order over the coupling constant $\chi$. Thus, they can be neglected
in Eqs.(\ref{Pi+-},\ref{Pir}) which are already written in the second
order over $\chi$.

Then Eq.~(\ref{D+-2}) leads to
\begin{widetext}
\begin{equation}\label{phkin2}
\frac{\partial N_\mathrm{ph}}{\partial t}
=  2 \pi\chi^2 \int d\varepsilon d\omega\rho_a(\varepsilon)
\rho_a(\varepsilon+\omega)\rho_\mathrm{ph}(\omega)\left[1-n_a(\varepsilon)\right]
n_a(\varepsilon+\omega)
- \int d \omega
     \gamma_{eh}(\omega)n_\mathrm{ph}(\omega)\rho_\mathrm{ph}(\omega),
    \end{equation}
where $\rho_\mathrm{ph}(\varepsilon)=(1/2\pi)\gamma_{eh}(\omega)/[(\omega-\Omega)^2+\gamma_{eh}^2(\omega)/4]$
is the vibrational density of states with the broadening, and the
vibrational damping rate (or linewidth) $\gamma_{eh}$ due to electron-hole pair excitation
is
\begin{equation}\label{ehpdamping}
\gamma_{eh}(\omega) = 2\pi \chi^2 \int d \varepsilon\left[ n_{a}(\varepsilon)
-n_{a}(\varepsilon+\omega)\right]\rho_{a}(\varepsilon)\rho_a(\varepsilon+\omega
)
\approx 2 \pi \chi^2 \frac{\Delta_{s}\rho^2_{a}(\mu_{s})+
\Delta_{t}\rho^2_{a}(\mu_{t})}{\Delta}\ \omega.
\end{equation}
\end{widetext}
The last approximate equation in~(\ref{ehpdamping}) is valid in the limit of low-temperature,
$k_BT\ll \omega$ and  slow-varying $\rho_a(\varepsilon)$ over $eV$.

Adding and subtracting integral term $\int d \omega
\gamma_{eh}(\omega)n_T(\omega)\rho_\mathrm{ph}(\omega)$ to the kinetic
equation~(\ref{phkin2}), it can be transferred to a more
transparent form, showing that in the absence of the applied
voltage the stationary phonon distribution function is the
equilibrium one, $n_T(\omega)=(e^{\omega/k_BT}-1)^{-1}$,
\begin{widetext}
\begin{equation}\label{kinEqua}
0=\frac{\partial N_\mathrm{ph}}{\partial t}= \int d \omega \rho_\mathrm{ph}(\omega)
\left\{
\Gamma_\mathrm{in}(\omega) - \gamma_{eh}(\omega) \left[n_\mathrm{ph}(\omega) - n_T(\omega)\right]
\right\}.
\end{equation}
Here
\begin{equation}\label{gammain}
\Gamma_\mathrm{in}(\omega)=2 \pi \chi^2 \frac{
\Delta_{s}\Delta_{t}}{\Delta^2}
\int d\varepsilon  \rho_{a}(\varepsilon) \rho_{a}(\varepsilon + \omega)
\left[ n_{s}(\varepsilon + \omega) -
n_{t}(\varepsilon + \omega) \right] \left[ n_{s}(\varepsilon) -
n_{t}(\varepsilon) \right],
\end{equation}
is the vibrational generation rate function.
\end{widetext}

\section{Rates of vibrational generation and adsorbate motion by tunneling current}
\label{sec:Phgenrate}

Kinetic equation Eq.~(\ref{kinEqua}) in the stationary regime allows
us to find the stationary
distribution function $n_\mathrm{ph}(\omega)$ which deviates
from the equilibrium one $n_T(\omega)$. The consequences of this deviation
on the rate of adsorbate motion can be analysed within
the truncated oscillator model, introduced and fully described in Ref.~\onlinecite{Gao97}.
The main quantities in the truncated oscillator model~\cite{Gao97}
are the the vibrational excitation and deexcitation rates between the nearest-neighboring
levels of the harmonic oscillator, $\Gamma_\uparrow$ and $\Gamma_\downarrow$, respectively.
Knowing $\Gamma_{\uparrow,\downarrow}$, it becomes possible, as shown in Ref.~\onlinecite{Gao97},
to introduce the adsorbate effective temperature
 \begin{equation}\label{teff}
 T_\mathrm{eff}=\frac{\Omega}{k_B}
 \left(
 \mathrm{ln}\frac{\Gamma_\downarrow}{\Gamma_\uparrow}
 \right)^{-1},
 \end{equation}
and, solving the oscillator master equation, to estimate the rate of
adsorbate motion as~\cite{Gao97}
\begin{equation}\label{rate1}
R(V)=n\Gamma_{\uparrow}(\frac{\Gamma_{\uparrow}}{\Gamma_{\downarrow}})^{n-1},
\end{equation}
where $n$ is the number of the vibrational levels within the
truncated harmonic approximation for the potential well, and
$(n-1) \Omega$ is assumed to be close to the barrier
height. Equation~(\ref{rate1}) can be also written~\cite{Gao97} as
Arrhenius-like expression via the effective temperature,
\begin{equation}\label{rate2}
R(V)=n\Gamma_{\uparrow}\exp [-\frac{(n-1) \Omega}{k_BT_\mathrm{eff}(\Omega)}].
\end{equation}

In what follows, we generalize the equations for the excitation and deexcitation
rates from Ref.~\onlinecite{Gao97}, in order to include the effect of the finite
lifetime broadening of the vibrational excitations.
The solution of the kinetic equation~(\ref{kinEqua}) for the
stationary vibrational occupation number
\begin{equation}\label{nphOmega}
  N_\mathrm{ph} = \int d \omega \rho_\mathrm{ph}(\omega)n_\mathrm{ph}(\omega)
\end{equation}
can be read as the \textit{ effective} Bose-Einstein distribution
function
\begin{equation}\label{nphOmega2}
N_\mathrm{ph} =
\left[
\exp (\Omega/k_BT_\mathrm{eff})-1
\right]^{-1}
\end{equation}
in the presence of the
IET-driven generation $\Gamma_\mathrm{iet}$ in competing to its damping
$\gamma$. Comparing the latter with the definition of the effective temperature
via the excitation and deexcitation rates, Eq.~(\ref{teff}), and
taking into account the explicit
form of the kinetic equation (\ref{kinEqua}), these rates can be written as
\begin{equation}\label{GammaUpDown}
  \Gamma_\uparrow=n_T(\Omega)\bar{\gamma}_{eh}+\Gamma_\mathrm{iet}, \,
\Gamma_\downarrow=[n_T(\Omega)+1]\bar{\gamma}_{eh}+\Gamma_\mathrm{iet},
\end{equation}
where, in contrast to Ref.~\onlinecite{Gao97}, the effect of the
vibrational lifetime broadening is included into
the generalized form of the IET vibration generation rate
via the integration of $\Gamma_\mathrm{in}$ over the lifetime-broadened
vibrational density function,
\begin{equation}
\label{GammaIET}
\Gamma_\mathrm{iet}= \int d \omega \rho_\mathrm{ph}(\omega)\Gamma_\mathrm{in}(\omega),
\end{equation}
and
\begin{equation}\label{bargamma}
 \bar{\gamma}_{eh} = \frac{
 \int d \omega \rho_\mathrm{ph}(\omega) \gamma_{eh}(\omega) n_T(\omega)
 }
 {
 n_T(\Omega)
 }
\end{equation}
is an averaged vibrational linewidth. Actually, because in reality the vibrational linewidth
is always relatively small, $\gamma_{eh} \ll \Omega$, and $\gamma_{eh}(\omega)$ is a
slowly changing function at $\omega \sim \Omega$, a good approximation
for Eq.~(\ref{bargamma}) is $\bar{\gamma}_{eh} \approx  \gamma_{eh}(\Omega)$.

We would like to note here that, in principle, we have to  include additively
(see the discussion above)
the relaxation rates
via other channels due to phonon excitations in the substrate
$\gamma_\mathrm{ph}$, if they are important, and $\gamma_{eh}$ is replaced by total
relaxation rate,
$\gamma=\gamma_{eh}+\gamma_\mathrm{ph}$.~\cite{Gao97}

\subsection{Physical meaning of $\Gamma_\mathrm{iet}$} \label{subsec:Giet}

Let us discuss the physical meaning of $\Gamma_\mathrm{iet}$ from the point of view of
the elementary processes of the elastic and inelastic electron tunneling through
the adsorbate-induced resonance.

\begin{figure}
\includegraphics[width=8cm]{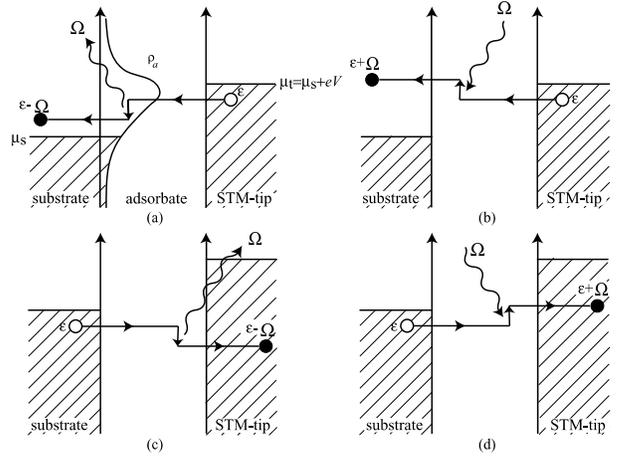}
\caption{
\label{fig:1}
Schematic representation of the inelastic electron transfers
through the adsorbate
from tip to
substrate with emission and absorption of phonons $R_{ts}^{+}$ and
$R_{ts}^{-}$,  (a) and (b), respectively.
Same
from substrate to tip
with emission and absorption of phonons $R_{st}^{+}$ and
$R_{st}^{-}$,  (c) and (d), respectively.
}
\end{figure}

As has been noted in Ref.~\onlinecite{Mii02}, the main difference between the
elementary processes composing the inelastic current and the vibrational generation rate,
is the fact that at $T\neq 0$ there are inverse (hole)
inelastic currents, which come with different signs to the total current and to
the vibrational generation rate. The expression for the inelastic tunneling
rate in the second order over $\chi$, [see Eqs.~(14) from Ref.~\onlinecite{Mii02}],
can be written as
\begin{eqnarray}
\label{Rine}
&& I_\mathrm{in}^{(2)}/e \equiv R_\mathrm{in}
\\ \nonumber
&=& 2 \pi \chi^2  \frac{\Delta_{s}\Delta_{t}}{\Delta^2}
\int d\varepsilon \int  d\omega
 \rho_{a}^{(0)}(\varepsilon) \rho_{a}^{(0)}(\varepsilon + \omega)  \rho_\mathrm{ph}(\omega)
\label{eq:in-a}
\\ \nonumber
&\times&
\left\{
- n_{s}(\varepsilon + \omega)\left[1 - n_{t}(\varepsilon)\right] \left[1 + n_\mathrm{ph}(\omega)\right] \right.
\\ \nonumber
&-& n_{s}(\varepsilon) \left[1 - n_{t}(\varepsilon + \omega)\right]n_\mathrm{ph}(\omega)
 \\  \nonumber
&+& \left[1 - n_{s}(\varepsilon)\right] n_{t}(\varepsilon + \omega) \left[1 + n_\mathrm{ph}(\omega)\right]
 \\  \nonumber
&+&\left[1 - n_{s}(\varepsilon + \omega)\right] n_{t}(\varepsilon)
n_\mathrm{ph}(\omega) \left.\right\}.\nonumber
 \end{eqnarray}
As schematically illustrated in Fig.~1  each term of this equation has its own
 inelastic tunneling process, and it can be deconvoluted into
\begin{equation}\label{Rinelec}
  R_\mathrm{in}=+R_{st}^{+} +R_{st}^{-}-R_{ts}^{+}-R_{ts}^{-}.
\end{equation}
Here $R_{st}^{\pm}$ denotes the electron transfer rate
through the adsorbate level
with vibrational emission/absorption  from substrate to tip
(or a hole from tip to substrate), and
$R_{ts}^{\pm}$ is the reverse process from tip to substrate.
In Figure~1, the initial electron states are shown as empty circles below
the corresponding Fermi levels. The final states are shown as
solid circles. In order that the partial currents $R_{ij}^{\pm}$ are
nonzero, the final states should be empty. For example, at $T = 0$ and the
applied voltage shown in Fig.~1, the currents shown in panels
(c) and (d) are zero. They become slightly nonzero at finite temperatures,
due to thermal excitation of holes below the Fermi levels.

We turn now to the calculation of the vibrational generation
rate. On the one side, it is zero in the stationary situation, Eq.~(\ref{kinEqua}).
On the other side, it
can be calculated within the same accuracy over $\chi$ as
Eq~(\ref{Rinelec}),
summing up the electron transfer rates with vibrational generation
and subtracting the rates with absorption, including the
processes of generation and absorption
leaving the electrons in substrate and in tip. The latter do not
contribute into current and, thus, are absent in Eq~(\ref{Rinelec}).
Then,
\begin{equation}\label{Rphfull}
 0 = R_\mathrm{ph}^{st}+R_\mathrm{ph}^{ss}+R_\mathrm{ph}^{tt},
\end{equation}
where, e.g.,
\begin{equation}\label{Rinph1}
R_\mathrm{ph}^{st} = R_{st}^{(+)} - R_{st}^{(-)}+R_{ts}^{(+)}- R_{ts}^{(-)},
\end{equation}
compare the signs here with that in Eq.~(\ref{Rinelec}). After some algebra we obtain,
\begin{eqnarray}
&& R_\mathrm{ph}^{st} \equiv R_\mathrm{ph}^{st}\{n_\mathrm{ph}(\omega)\} \label{operator}\\ \nonumber
&=& 2 \pi \chi^2 \frac{\Delta_s \Delta_t}{\Delta^2}\int d\varepsilon \int  d\omega
 \rho_{a}^{(0)}(\varepsilon) \rho_{a}^{(0)}(\varepsilon + \omega)
  \rho_\mathrm{ph}(\omega)
  \nonumber \\
 &\times& (
 n_s(\varepsilon+\omega)\left[1-n_t(\varepsilon)\right] +
 n_t(\varepsilon+\omega)\left[1-n_s(\varepsilon)\right]
 \label{Rinph2}\\ \nonumber
&+& n_\mathrm{ph}(\omega)
\left[ n_s(\varepsilon+\omega)+ n_t(\varepsilon+\omega)
-n_s(\varepsilon) -n_t(\varepsilon)\right]).
\end{eqnarray}
It appears that, in order to calculate $\Gamma_\mathrm{iet}$, Eq.(\ref{GammaIET}),
 via  $R_\mathrm{ph}^{st}$,
we have to replace in  Eq.~(\ref{Rinph2}) the stationary nonequilibrium
distribution function  $n_\mathrm{ph}(\omega)$ by the equilibrium one.
Using notation of Eq.~(\ref{operator}), it can be written as
\begin{equation}\label{Giet2}
  \Gamma_\mathrm{iet} =  R_\mathrm{ph}^{st}\{n_T(\omega)\}.
\end{equation}
This means that there is a small but systematic difference between
$ \Gamma_\mathrm{iet}$ Eqs.~(\ref{Giet2},\ref{Rinph1})  and
absolute value of IET rate  $|R_\mathrm{in}|$ Eqs.~(\ref{Rine},\ref{Rinelec}),
growing with the increase
of deviation of stationary vibration population $n_\mathrm{ph}$ from the
equilibrium one $n_T$, see also Eq.~(\ref{Rine2}) below.

\subsection{Limit of $T=0$ and no vibrational broadening} \label{subsec:Teq0}

Before presenting the results of numerical calculation of
$\Gamma_\mathrm{iet}, R_\mathrm{in}$, and $R$ at finite temperatures
and including the effect of vibrational broadening, it is
helpful to analyze the situation in
the limit $|eV|\gg  \Omega \gg k_BT$, neglecting the vibrational
broadening, $\rho_\mathrm{ph}(\omega) =
\delta(\omega-\Omega)$. For simplicity, we will also assume
$\Delta_{s} \gg \Delta_{t}$. We expect to obtain the results equivalent to those
in Gao \textit{et al}.~\cite{Gao97}

Equation~(\ref{nphOmega}) then reduces to
\begin{equation}\label{Nph}
N_\mathrm{ph} = n_\mathrm{ph}(\Omega) = n_T(\Omega) + \frac{\Gamma_\mathrm{iet}}{\gamma_{eh}(\Omega)},
\end{equation}
with
\begin{eqnarray}
 \label{GietAPP}
\Gamma_\mathrm{iet} &= & \Gamma_\mathrm{in}(\Omega) =
2\pi \chi^2\frac{\Delta_{s}\Delta_{t}}{\Delta^2}
\rho_{a}(\mu_{t})\rho_{a}(\mu_{t}-\eta\Omega)
\frac{F(x)}{ \Omega}
\nonumber \\
&\simeq&
\frac{\Delta_t}{\Delta_s}\gamma_{eh}(\Omega)\frac{F(x)}{ \Omega},
\end{eqnarray}
where $F(x)=x\theta(x), \quad x=|eV|- \Omega$, and
$\eta={\rm sign}(V)$ is the sign function and $\theta$ is the step function.
In  the last step we have used $\rho_{a}(\mu_{t}-\eta\Omega)\simeq\rho_{a}(\mu_{t})$.
In the limit $|eV| \gg \Omega $
 Eq.~(\ref{GietAPP})  becomes a linear function of applied
 voltage, in agrement with the result obtained by
 Gao \textit{ et al.}~\cite{Gao97} from a Boltzmann
 distribution among different vibrational levels.

\begin{figure}
\includegraphics[width=8cm]{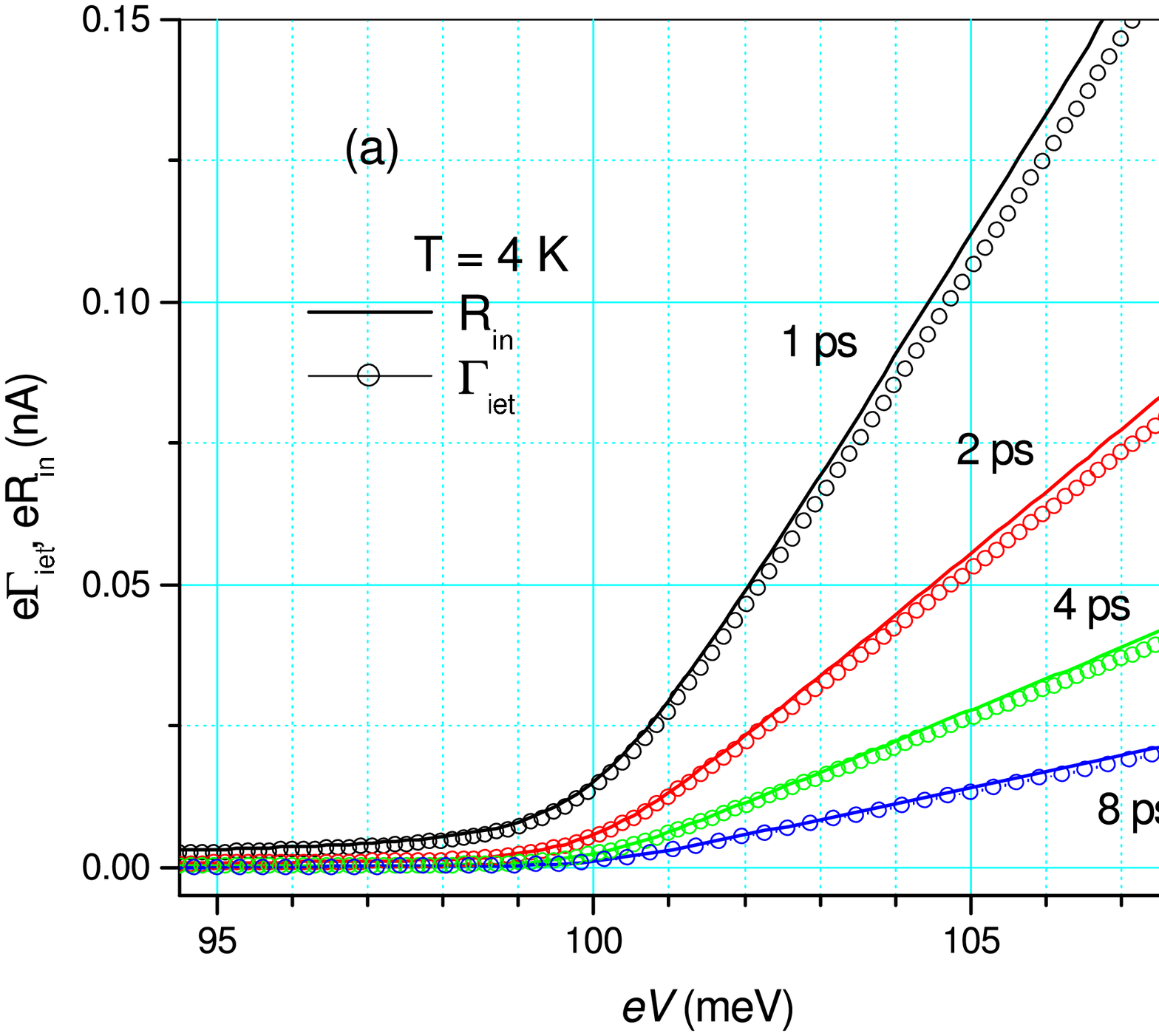}\\
\includegraphics[width=8cm]{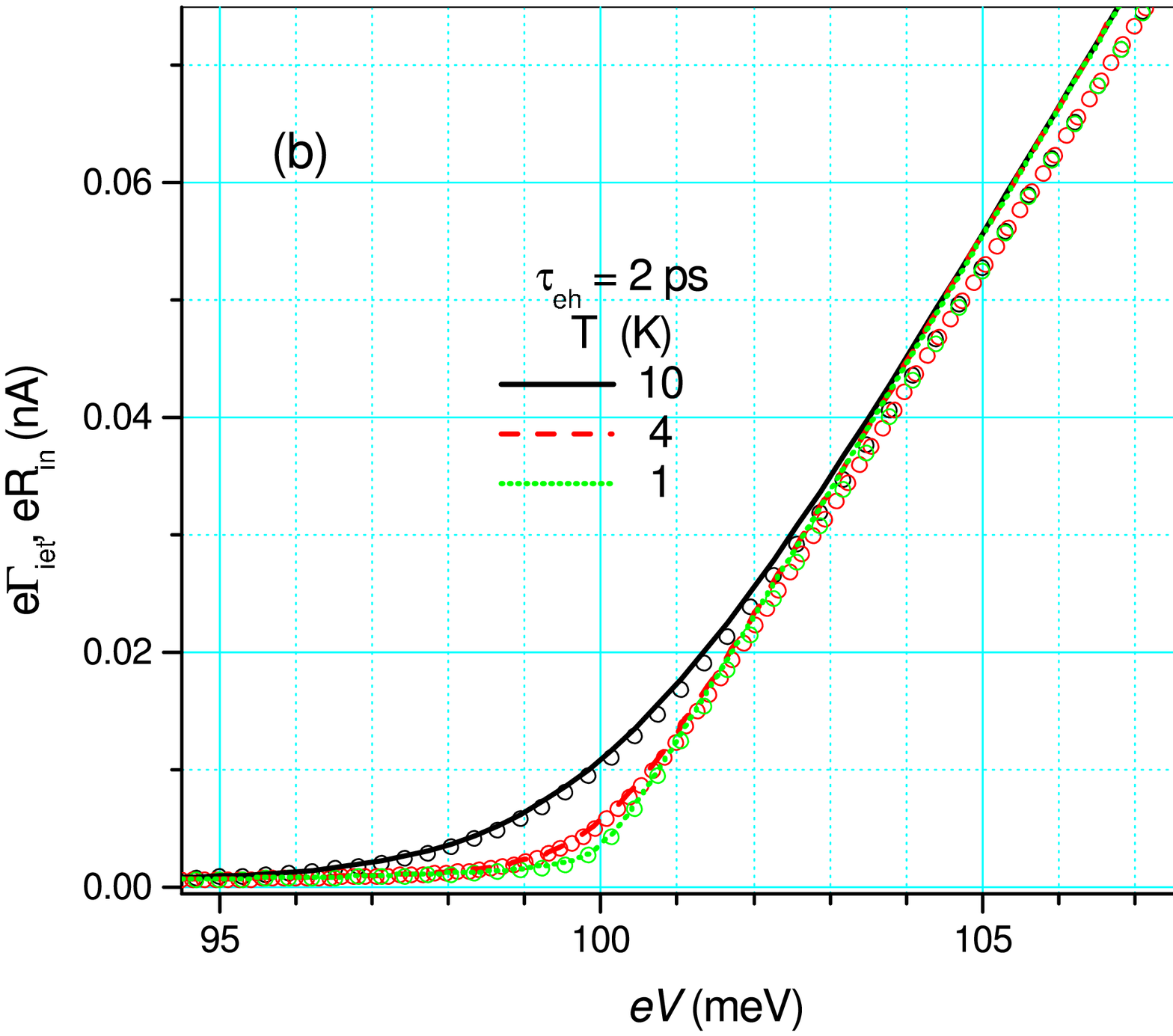}
\caption{Voltage dependences of the
inelastic current $R_\mathrm{in}$ and
vibrational generation rate
$\Gamma_\mathrm{iet}$  (lines and lines with circles,
respectively), calculated for $T = 4$~K and different vibrational lifetimes (a);
for $\tau_{eh} = 2$~ps and different temperatures (b).
The vibrational frequency is taken $\Omega = 100$~meV.
See other parameters in the text.
\label{fig:2}}
\end{figure}

\begin{figure}
\includegraphics[width=8cm]{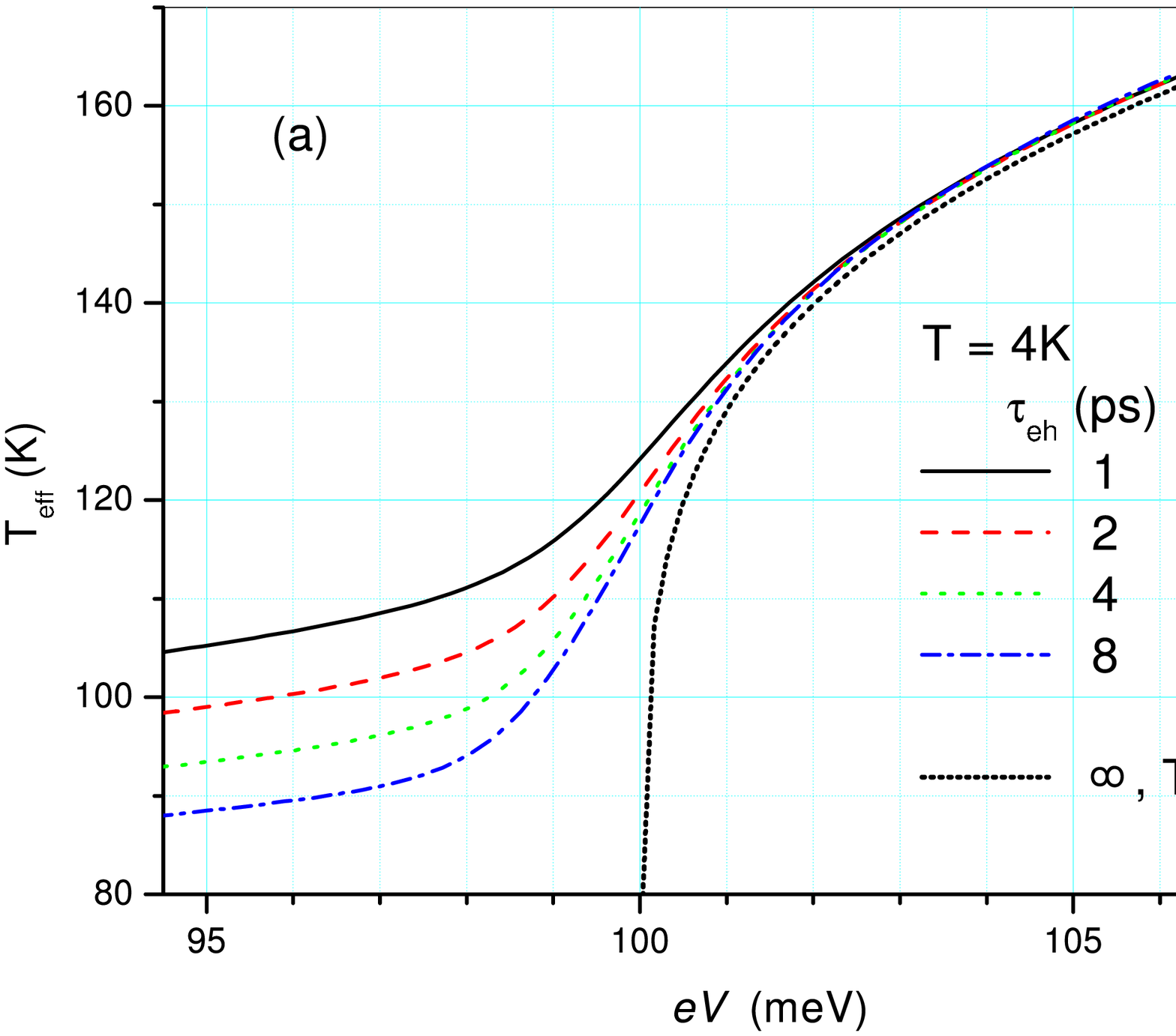}\\
\includegraphics[width=8cm]{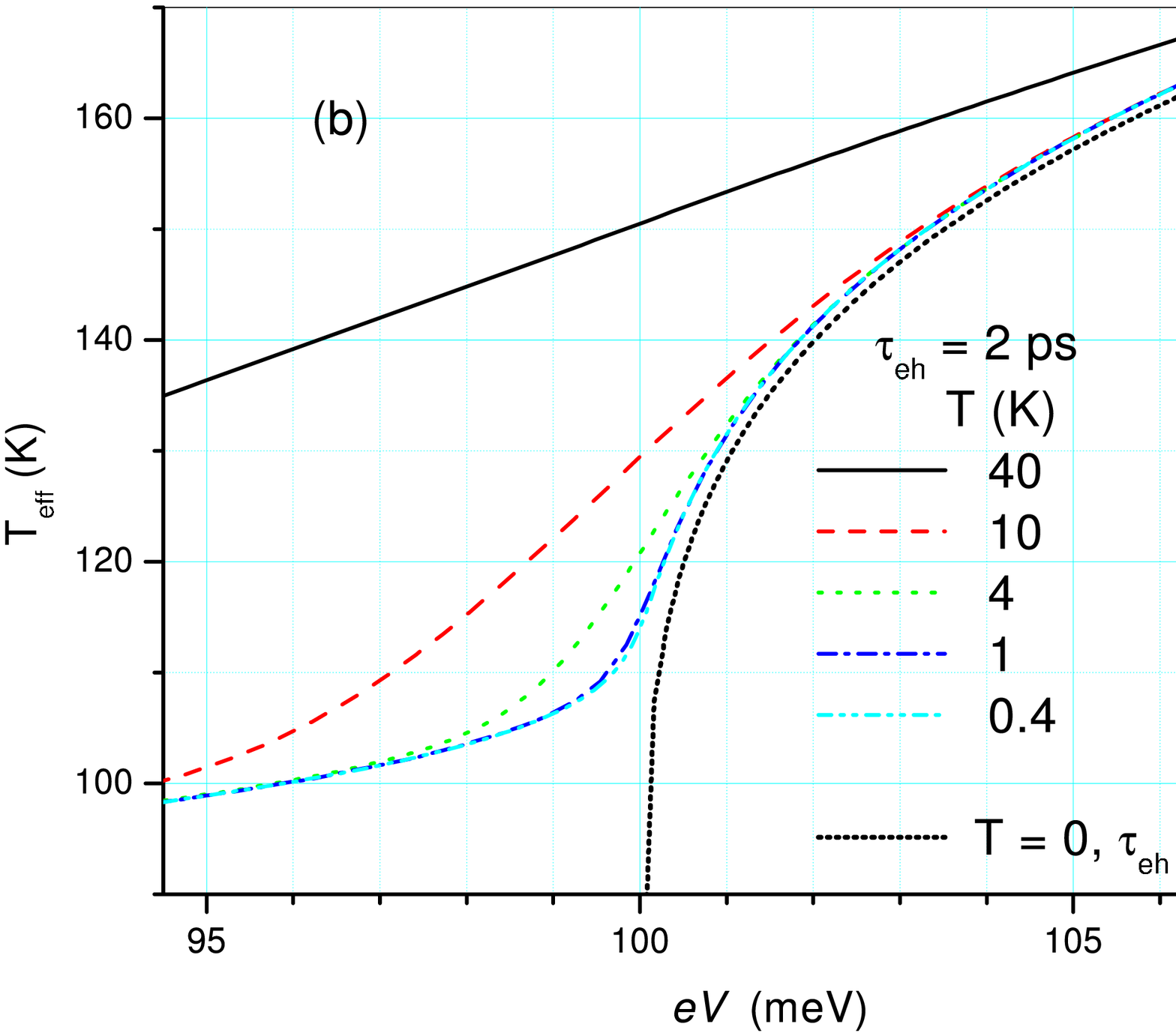}
\caption{Voltage dependences of the effective temperature,
calculated for bath temperature $T = 4$~K and different vibrational
lifetimes (a); for $\tau_{eh} = 2$~ps and different temperatures (b).
Short dotted line in both panels
shows the effective temperature in the
simple approximation neglecting the phonon broadening effect, $\tau_{eh} = \infty $,
and at $T = 0$, $\Gamma_\mathrm{iet}(V) \propto |eV|-\Omega$.
\label{fig:3}
}
\end{figure}

As to a relation between $R_\mathrm{in}$ and $\Gamma_\mathrm{iet}$ in this
limit,
comparing Eqs.~(\ref{Rine}) and (\ref{Giet2}), we find that
\begin{equation}\label{Rine2}
|R_{\mathrm{in}}|-\Gamma_\mathrm{iet}
=2\left(\frac{\Delta_t}{\Delta_s}\right)^2 \gamma_{eh}^s \frac{(|eV|-\Omega)|eV|}{\Omega^2}\Theta(|eV|-\Omega).
\end{equation}
Because $\left(\Delta_t/\Delta_s\right)^2 \ll 1$,
$|R_\mathrm{in}|\approx\Gamma_{iet}$.
The linear
relation between  $\Gamma_\mathrm{iet}$
(or approximately $R_\mathrm{in}$)
and the bias voltage
holds for $eV \gg \Omega$, which is a case of a Xe atom
transfer between a Ni substrate and a STM tip ($ \Omega$=4 meV
and $V$=20-200~mV).~\cite{Eigler91,Gao97}

As to the adsorbate motion rate, in this limit
a linear dependence of $\Gamma_\mathrm{iet}(V)$ on $|eV| - \Omega$
(Eq.~(\ref{GietAPP})) and a condition of
$\gamma\gg\Gamma_\mathrm{iet}(V)$ yield the power-law dependence
\begin{equation}\label{rate4}
R(V)\propto (|eV|- \Omega)^n
\end{equation}
 for Eq.~(\ref{rate1}), in agreement
with the analysis in Ref.~\onlinecite{Gao97}. The opposite
case $\gamma\ll\Gamma_\mathrm{iet}(V)$ implies infinite effective temperature  which is not
consistent with an assumption of activation-type-law for a reaction rate Eq.(\ref{rate2}).

\begin{figure}
\includegraphics[width=8cm]{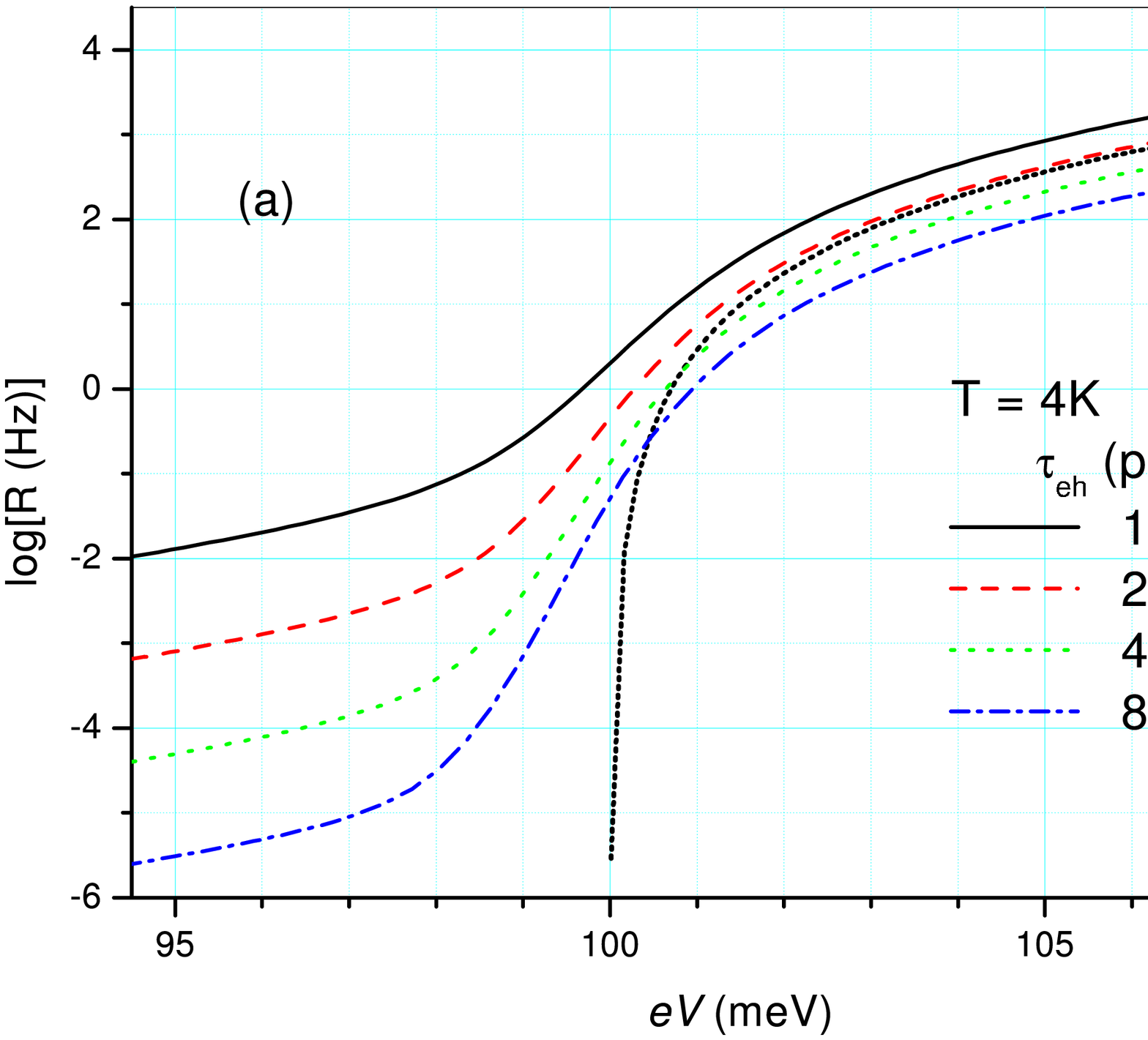} \\
\includegraphics[width=8cm]{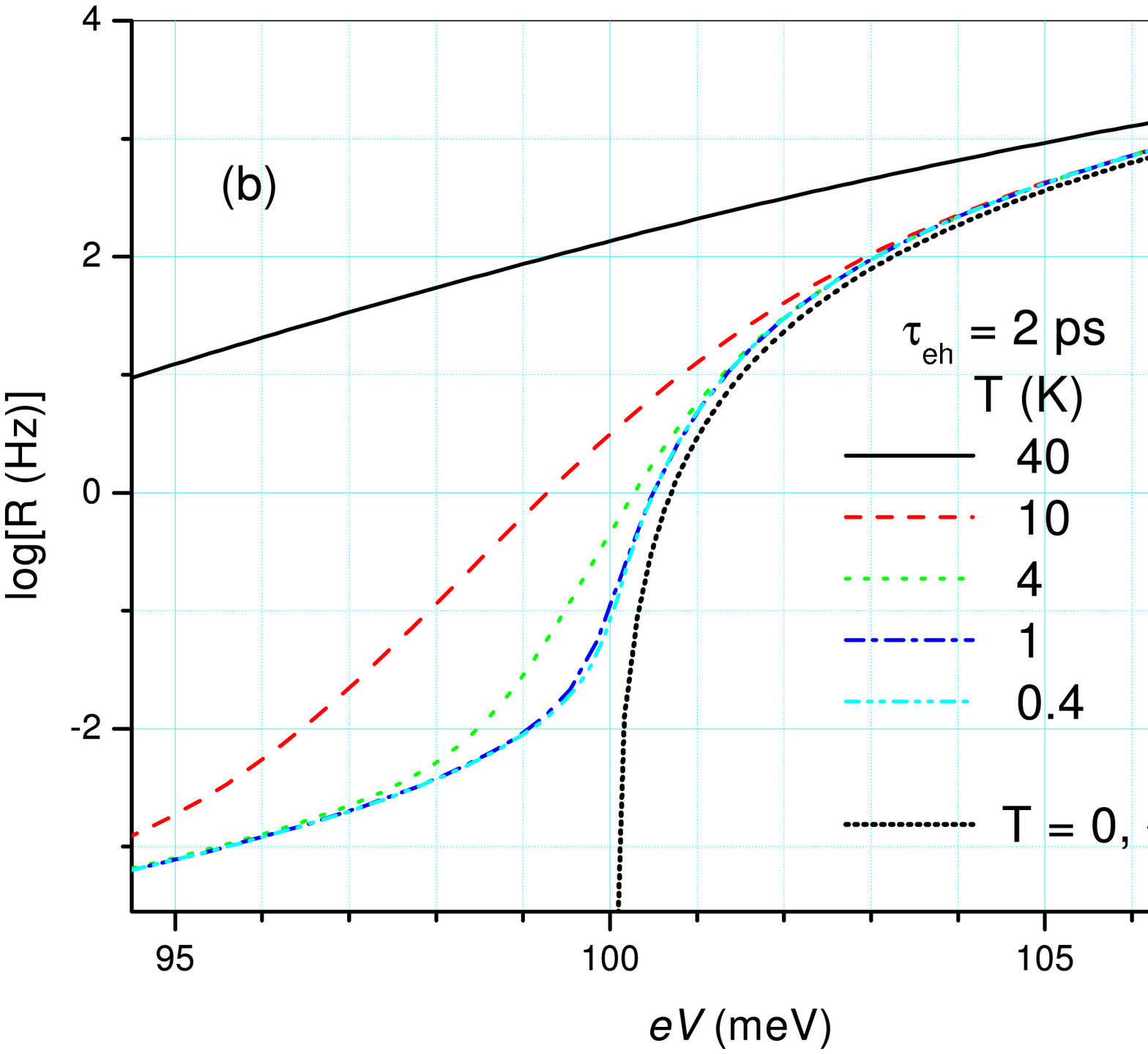}
\caption{Voltage dependences of the logarithm of the transfer rate,
calculated for bath temperature $T = 4$~K and different vibrational
lifetimes (a); for $\tau_{eh} = 2$~ps and different temperatures (b).
Short dotted line in both panels
shows the transfer rate in the
simple approximation neglecting the phonon broadening effect, $\tau_{eh} = \infty $,
and at $T = 0$, $R(V) \propto (|eV|-\Omega)^n$, $n = 3$.
\label{fig:4}
}
\end{figure}

\subsection{Numerical examples}\label{subsec:Num}

We are in a position to compare the relation between inelastic
tunneling rate $R_{in,e}$ and vibrational generation rate
$\Gamma_\mathrm{iet}$ as a function of $V$ for several values of the
vibrational relaxation time and temperature. Numerical calculations are performed
for a set of parameters: $\varepsilon_a$=2 eV above the Fermi
level of the substrate, $\Delta_s$= 1.0 eV, $\Delta_t$=0.025 eV,
$ \Omega=$100 meV, and different temperatures $T = 0.4, 1, 4$, and 40~K.
The electron-vibrational coupling constants were taken
$\chi = 500, 354, 250$ and 177~meV, which at $eV = 0$ and given $\Delta_{s,t}$
correspond to vibrational
linewidth $\gamma_{eh} = 0.66,$ 0.33, 0.16, and 0.08~meV
(or vibrational lifetimes $\tau_{eh} =  \hbar/\gamma_{eh}= $1, 2, 4, and 8~ps),
respectively.

Figure~\ref{fig:2} shows the
calculated vibrational generation rates Eq.~(\ref{GammaIET}) (circles) and
the corresponding inelastic currents Eq.(\ref{Rine}) (lines)
at fixed temperature for different vibrational lifetimes (panel a) and
for fixed vibrational lifetime for different temperatures (panel b).
The generation rates are given in nA and also in $\mu$eV (left and right vertical axes
in Fig.~\ref{fig:2}). Because the vibrational linewidths are taken in the range of
0.088 - 0.66~meV, it is clear that all the calculated rates are well within
the condition $\gamma\gg\Gamma_\mathrm{iet}(V)$.
It is seen that, in agreement with the analysis above, the inelastic current
starts to exceed slightly the vibrational generation rate above the threshold
of inelastic tunneling. As to the latter, it grows approximately linearly,
$\propto |eV| - \Omega$ above the threshold. The differences from the simple
linear dependence arise near the threshold. Instead of abrupt switching on of the
inelastic current and vibrational generation rate, there is a gradual growing of both.
The physical reason is the temperature and lifetime broadening of the IET threshold.
The temperature broadening has been analyzed previously (see, e.g., in
Ref.\onlinecite{Gao94}).
The lifetime broadening was not analyzed, at least in part of its influence on the overheating and
adsorbate motions. As we will see below, a different influence of
temperature and lifetime broadenings allows, in principle, to separate
these two mechanisms from the voltage dependence of the overheating and of motion rate
below the IET threshold.

It becomes more evident from Fig.~\ref{fig:3}, where the calculated voltage dependencies of
the effective temperature Eq.~(\ref{teff}) for the same parameters as in
Fig.~\ref{fig:2} are displayed.

Note that above the IET
threshold, $|eV| \gg \Omega$
the overheating does not depend on the vibrational lifetime as well as on
the temperature and approaches a universal linear dependence Eq.~(\ref{GietAPP})
$ \propto |eV| - \Omega$.
The reason for this is the cancellation of $\chi^2$ in  the nominator and denominator
of $\Gamma_\mathrm{iet}/\gamma_{eh}$, if the e-h scattering Eq.~(\ref{ehpdamping})
is the main channel of vibrational relaxation, $\gamma \approx \gamma_{eh}$.
As a result, the stationary vibrational occupation number (\ref{Nph}),
effective temperature (\ref{teff}), and rate of adsorbate motion (\ref{rate1})
do not depend on the  adsorbate-vibration coupling constant.
 We have already mentioned
this fact previously.~\cite{Mii02} This effect takes place only above the
IET threshold.

Whereas near and below the threshold, $|eV| \lesssim \Omega$, the effective
temperature becomes lifetime and bath temperature dependent.
First of all note that, in a seeming contradiction with a simple
understanding that a shorter lifetime means better cooling, an
opposite dependence is seen in Fig.~\ref{fig:3}a: the overheating \textit{grows}
with the lifetime decrease! The explanation is very simple: at low
temperatures and below the
IET threshold, the only reason of the nonzero inelastic current and
overheating is the finite lifetime broadening. Which grows with the
lifetime decrease.
Mathematically, an additional power of $\chi^2$ is brought into
$\Gamma_\mathrm{iet}$ below the IET threshold because of integration
of $\Gamma_\mathrm{in}(\omega) \propto \chi^2$ in Eq.~(\ref{GammaIET})
over the tail of the vibrational density of states at large detuning
$\gamma_\mathrm{ph}(\omega) \propto \gamma_{eh}/(\omega-\Omega)^2$,
which is $\propto \chi^2$ too.

Figure~\ref{fig:3}b shows that, as anticipated, the temperature broadening is
negligibly small in comparison with lifetime broadening, for lower temperatures
$k_B T \lesssim \gamma_{eh}$. For example, all $T_\mathrm{eff}$ curves in
Fig.~\ref{fig:3}b, calculated with $\gamma_{eh} = 0.33$~meV, merge
into a single limiting curve starting from bath temperatures $T < 1$~K
($k_B T < 0.9$~meV). This limiting curve corresponds to a $T=0$ solution
for effective temperature with a finite vibrational lifetime.

Another interesting thing can be seen
in Fig.~\ref{fig:3}b: the temperature broadening at higher temperatures,
$k_B T \gtrsim \gamma_{eh}$, is localised
within a finite range of several $k_B T$ near the IET threshold,
whereas the finite lifetime broadening is characterized by a long tail of
strong overheating even at $|eV| \ll \Omega$. The explanation is that the latter is due
to Lorenzian tail of the vibrational density of states, whereas the temperature
broadening has a Gaussian tail.

Figure~\ref{fig:4} plots $R(V)$ Eq.(\ref{rate2}) for the same set of parameters
as before
and barrier hight equal to $n = 3$ vibrational excitation quanta.
It shows that although with the increase of voltage above the IET threshold
$eV > \Omega$, the rates of motion reach eventually the limiting
power law dependence Eq.(\ref{rate4}), in the vicinity of the threshold
this dependence is different, depending on the relative values
of $k_BT$ and $\gamma_{eh}$. And below the threshold, at $|eV| \ll \Omega - k_B T$,
the rate of motion is fully due to the vibrational lifetime
broadening, thus making it to grow with the decrease of
the vibrational lifetime.

\section{CONCLUDING REMARKS}
We have studied a formal relation between
the inelastic tunneling
current, vibrational generation rate and vibrational heating
caused by the current-induced vibrational excitation of single adsorbates on metal
surfaces. The nonequilibrium process of
the electron tunneling between the tip and substrate through the
adsorbate level is described by coupled Dyson and kinetic
equations of the Keldysh-Green's functions. The inelastic tunneling
current and the vibrational generation rate are obtained
within the second-order perturbation with respect to the
electron-vibration coupling based on the adsorbate-induced
resonance model.
We show
that the vibrational
generation rate slightly differs from the absolute value of the
inelastic tunneling rate.
The linear dependence of
the vibrational generation rate on the bias voltage is valid for the
extreme cases of $|eV|\gg \Omega$ and the slowly varying
adsorbate density of states over the applied voltage. Under these
conditions the rate of adsorbate motions via the step-wise
vibrational ladder climbing due to the inelastic tunneling current
is characterized by a simple power-law dependence on the bias
voltage or the tunneling current.
It is
found that the effects of bath temperature and vibrational damping
manifest themselves in
the vicinity and below the IET threshold, at $|eV|\lesssim \Omega$.
The temperature broadening manifests itself only
at relatively large bath temperatures, $k_B T \gtrsim \gamma_{eh}$.
The important thing is that whereas the bath temperature
broadening is confined to a narrow range of several $k_B T$,
the vibrational lifetime broadening effect extends to a much
larger interval of voltages below the IET threshold.

\textbf{ACKNOWLEDMENTS}
We would like to thank T. Mii for useful discussions.
This work was financially supported in part by the Russian Foundation
for Basic Research and the Russian Ministry of Science. S.T. thanks
Japan Society for the Promotion of Science fellowship program (JSPS)
(No. S02227) for supporting his visit to Toyama University. H.
U. is supported by Grant-in-Aid for Basic Research (B)
(No. 15310071) from JSPS.


\begin{thebibliography}{99}
\bibitem{footnote}
If the electron-hole pair scattering is not the only mechanism of the phonon
relaxation, other mechanisms should be included into $\Pi^r$.
\end{thebibliography}
\end{document}